\documentclass[pdflatex,sn-mathphys-num]{sn-jnl}

\usepackage{graphicx}%
\usepackage{multirow}%
\usepackage{amsmath,amssymb,amsfonts}%
\usepackage{amsthm}%
\usepackage{mathrsfs}%
\usepackage[title]{appendix}%
\usepackage{xcolor}%
\usepackage{textcomp}%
\usepackage{manyfoot}%
\usepackage{booktabs}%
\usepackage{algorithm}%
\usepackage{algorithmicx}%
\usepackage{algpseudocode}%
\usepackage{listings}%
\usepackage{geometry}
\begin{document}

\title[Article Title]{A Space-Time Fluid (Unabridged)}

\author*[1]{\fnm{Albert} \sur{Stebbins}}\email{stebbins@fnal.gov}

\affil*[1]{\orgdiv{Theoretical Astrophysics}, \orgname{Fermi National Accelerator Laboratory}, \orgaddress{\street{Box 500}, \city{Batavia}, \postcode{60510}, \state{IL}, \country{USA}}}

 \abstract{
 
\textbf{Purpose:} This essay is a retelling of general relativity in a language in which space-time geometry is expressed as a fluid.  This trivial and useful reformulation gives 1) a non-perturbative covariant description of cosmological inhomogeneities and 2) a simple formula describing how cosmic inhomogeneities are generated on super-horizon scales.
 
\textbf{Methods:} Equating the Ricci curvature with the associated matter stress-energy  gives a description of space-time geometry in terms of fluid properties.  These locally measurable (covariant) non-perturbative quantities are in some ways superior to commonly used "gauge invariant" quantities.  The dynamics of a quantity (kurvature) which describes cosmological inhomogeneities is described in detail. A detailed comparison is made of space-time fluid dynamics with that of a classical (Newtonian physics) fluid.

\textbf{Results:} The fluid lexicon permits an unambiguous definition of the velocity of space-time.  The evolution of the space-time fluid is in many ways identical with that of the classical fluid when expressed on Lagrangian coordinates.  Kurvature is a measure of the specific binding energy of the fluid and is a most useful covariant measure of cosmological inhomogeneities. For plausible matter models kurvature will increase, even on super-horizon scales, due to non-linear hydrodynamic effects rather than gravity.  This phenomena is also exhibited by classical fluids.

\textbf{Conclusion:} The space-time fluid representation of geometrodynamics gives a simple and useful description of the evolution of cosmological inhomogeneities. 

}




\maketitle

\section{Frequently Asked Questions}
\label{sec:FAQ}

The goal of this essay is to address a couple of questions regarding gravity and cosmology which the author has asked himself over the years.  These questions are in the context of normal gravity (Einstein's general relativity) and are as much questions about descriptions as they are about phenomenology.  

\subsection{Does Space Move?}

One finds in textbooks / online / in lectures statements that the "universe expands". This is often illustrated by something like the "raisin bread model": the universe is like raisin bread the space is like the dough which expands because of leavening while the galaxies like the raisins are carried along with the dough.  Such statements always seemed to me (and many others) to be an arcane description of a phenomenon more simply understood by anyone familiar with elementary mechanics: the galaxies are moving away from each other. In this simpler description the space is not dynamical and there is no sense in which it moves.  Of course we know that in metric theories of gravity (general relativity or GR) that space-time is dynamical (geometrodynamics).  Still that does not mean one can make sense of space itself moving.  Yet there are phenomena in GR like frame dragging which is as if space is moving in some quantifiable sense.  If there were a formal way of assigning a velocity to space that might prove useful in understanding geometrodynamics. 

\subsection{Generation of Super-Horizon Density Inhomogeneities?}
\label{InhomogeneityGenerationQ}

One can make the argument that the nearly isotropic temperature of the cosmic microwave background (CMB) along with the large angle correlations of the CMB temperature anisotropies shows that all visible parts of our universe are in causal contact with each other requiring a period of "inflation".  The large angle anisotropies, believed to reflect large scale gravitational potential inhomogeneities, indicates that density inhomogeneities have a power spectrum close to the Harrison-Zel'dovich form: $P_\rho[k]\propto k^{n_\rho}$ with $n_\rho\approx1$.  In flat space the evolution of the density field is governed by energy momentum conservation:
\begin{eqnarray}
\dot{\rho}&=&-\vec{\nabla}\cdot\vec{S}
\qquad
\dot{\vec{S}}=-\vec{\nabla}\cdot\overleftrightarrow{p} 
\nonumber \\
\rho[t,\vec{x}]&=&\rho[t_\mathrm{i},\vec{x}]
+\dot{\rho}[t_\mathrm{i},\vec{x}]\,(t-t_\mathrm{i})
+\int_{t_\mathrm{i}}^t dt'\,(t-t')\,
\vec{\nabla}\cdot(\vec{\nabla}\cdot\overleftrightarrow{p})[t',\vec{x}])
\label{rhofromp}
\end{eqnarray}
where $\rho$ is the density, $\vec{S}$ the energy flux and $\overleftrightarrow{p}$ is the stress tensor.  If there is no initial large scale power then, since the density is the gradient of the gradient of integrated pressure inhomogeneities, the spectral index of density is related to that of pressure by $n_\rho=n_\mathrm{p}+2\times2$. If the pressure has no large scale correlations, $n_\mathrm{p}\ge0$, then $n_\rho\ge4$ which is not what is observed. By this reasoning something must be initially correlated on large scales or alternately small scale correlations must be blown up to large scales. Inflation does that.

However this reasoning rests on flat space global energy-momentum conservation which we know does not hold in GR.  Nevertheless a similar argument can be made in the linear theory of cosmological inhomogeneities \cite{ZELDOVICH1965,1974A&A....32..197P,1983ApJ...268....1C,1986ApJ...302...39A,Veeraraghavan1990,PhysRevD.68.103515}.  In non-linear GR this reasoning will fail at some level but inhomogeneities are initially so small that these non-linear effects might be negligible.  On the other hand adding even a small source term to eq.~\ref{rhofromp}, even if it is not correlated on large scales, would add white noise, $n_\rho\approx0$, to $P_\rho$.  On the largest scales this would come to dominate the observed $n_\rho\approx1$ spectrum. Answering this question is thus important for understanding our universe!

As it turns out finding an answer to the first question helps to formulate an answer to the second.

\section{Semantic Solution}
\label{CenterOfMomentum}

For space to expand or contract one would require space to have a 4-velocity to either diverge or converge.  Geometrodynamics\footnote{The results presented here are a simple retelling of fundamentals of hydrodynamics/GR such as can be found in textbooks, e.g. \cite{Landau1987Fluid,Hawking:1973uf}.} describes the evolution of the curvature tensor(s) of a space-time so if space has a velocity one would want this velocity to derive from the curvature. One does not usually associate a rank 1 vector with these even ranked tensors. An exception occurs when the space-time has a continuous isometry and an associated Killing vector.  Killing vectors define conserved currents, an energy density current in the case of a time-like Killing vector, however cosmological inhomogeneities will not generally have isometries or their associated Killing vectors.

There are other ways to define vectors from even ranked tensors.  One of the simplest are the eigenvectors of these tensors.  Therefore a candidate {\it space-time 4-velocity} is the time-like future directed eigenvector of the Einstein curvature tensor\footnote{The eigenvectors of the Einstein and Ricci tensors are the same. N.B.Here I use a $\{-,+,+,+\}$ signature metric and occasionally $c=1$.}:
\begin{equation}
{G^\alpha}_\beta\,\bar{u}^\beta=-8\pi\,G\,\bar{\rho}\ \bar{u}^\alpha \ .
\end{equation}
where $\bar{\rho}$ is the matter density in the center-of-momentum frame and $\bar{u}^\alpha$ when properly normalized ($\bar{u}^\alpha\bar{u}_\alpha=-1$) is the 4-velocity of this center-of-momentum frame.\footnote{Defining space-time geometry in terms of a locally defined 4-velocity field is the covariant approach to cosmological perturbations developed in \cite{Ellis:1994sm} and references therein.  Their choice velocity is more observationally relevant then the center-of-momentum velocity.}  In the raisin bread model this amounts to saying that the dough follows the raisins. This seems too trivial to be useful.

Objections to using $\bar{u}^\alpha$ for the velocity of space-time include
\begin{itemize}
    \item $\bar{u}^\alpha$ is already the velocity of the matter!
    \begin{itemize}
        \item there is no value added!
    \end{itemize}
    \item it seems silly to peg the space-time velocity to the center-of-momentum of any tenuous bit of matter that happens to be present in a region. 
    \begin{itemize}
        \item for example the gas surrounding a star can and does move in any direction. A star's more rigid tidal field seems a better description of the "gravity" around the star.   
    \end{itemize}
    \item there is no unique $\bar{u}^\alpha$ wherever $G^{\alpha\beta}$  ($R^{\alpha\beta}$) has Lorentz symmetries.\footnote{N.B. $\bar{u}^\alpha$ is well defined except for a set of measure zero of $T_{\alpha\beta}$'s where either one of the its eigenvectors is a null vector or where one or more of the pressure eigenvalues is $\pm$ the density eigenvalue. This excludes vacuum $T_{\alpha\beta}=0$, a pure cosmological constant $T_{\alpha\beta}=-\Lambda\,g_{\alpha\beta}$, a pure non-interacting scalar fields with space-like gradient, idealized laser beams, etc.  It does not usually include $T_{\alpha\beta}$'s which violate the various energy conditions, such as phantom dark energy.}
    \begin{itemize}
        \item this is common in "important" cases, for example empty space or one containing only a cosmological constant.
        \item what about the empty space between atoms?  Isn't that everywhere?
    \end{itemize}
\end{itemize}
I sympathize with many of these objections but nevertheless will propose conflating matter and space-time properties, at least as a conceptual exercise. The center-of-momentum velocity is already a property of gravitational system so a new lexicon will not change the phenomenology.  What one chooses to call $\bar{u}^\alpha$ is, of course, only a question of semantics.  Semantics can help or hinder our understanding of things.

One reason it sounds strange to think of $\bar{u}^\alpha$ as the space-time velocity is that we think of matter and gravity as two distinct things.  There is no good reason for this at least in the context of GR.  We say that curvature is a property of space-time, density is a property of matter, charge density is a property of the electron spinor field, etc.  We don't say {\it "electrons cause space to charge"} yet we do say {\it "matter causes space to curve"} as if they are separable entities.  If one takes Einstein's equations to mean that space-time curvature is a property of matter then {\it "matter has curvature"} it doesn't cause curvature. If the two are part and parcel of the same thing then it would seem natural to assign the matter velocity to the space-time.  Space-time cannot move independently of matter any more than charge moves independently of charged particles.

This essay henceforth uses a lexicon\footnote{
I am not claiming that this sort of association of words and concepts is new.}
where space-time, meaning the Ricci curvature, is said to possess all the properties of a fluid: e.g.: fluid elements, velocity, $\bar{u}^\alpha$, streamlines, shear, $\bar{\sigma}_{\alpha\beta}$, vorticity, $\bar{\omega}_{\alpha\beta}$, viscosity, $\bar{\mu}$, density, $\bar{\rho}$, pressure, $\bar{p}$ and anisotropic stress, $\bar{q}^{\alpha\beta}$. Since Ricci curvature is a property of both matter and space-time one could equally well refer to this fluid as the matter fluid or the {\it space-time fluid}.  I choose the latter to emphasize the geometrical aspects and also to minimize confusion with any number of "real" fluids of which the matter might consist. The $8\pi\,G$ in Einstein's equation just serves as a unit conversion factor much like $c$ converts between space and time or $c^2$ between pressure and density.

This lexicon can be used to describe space-times containing any type of matter: gasses, liquids or solids, scalar or electro-magnetic fields, etc.  In addition since space-time fluid properties refer to space-time geometry one can use the same lexicon to describe geometry in modified theories of gravity where the relation between matter and curvature differs from Einstein's equation.

For $\bar{u}^\alpha$ to be uniquely defined $G^{\alpha\beta}$ must have no Lorentz (boost) symmetries otherwise the time-like eigenvector is not uniquely defined.  While one could choose among degenerate eigenvectors it is better to concede that space-time has no velocity when this occurs just as matter has no velocity where there is no matter.  Empty space then has no defined velocity\footnote{
My perspective on the empty space between particles is that matter is made of fields (waves) not particles and this does not happen.  Usually matter fills all of space.  However adapting this formalism to quantum fields is problematic}.
Since much of GR phenomenology is related to isolated objects in vacuum this is a real limitation.  However the space-time fluid velocity is defined whenever and wherever 
"normal\footnote{
By normal I mean $\mathrm{max}[|\bar{p}_i|]<|\bar{\rho}|\,c^2$ where $\bar{p}_i$ are the eigenvalues of the pressure tensor},
matter" is present and even for most abnormal matter. Normal matter fills space for nearly all of cosmology and in many other situations.  
The space-time fluid tends to avoid singularity because {\it it is} matter which pushes back against compression. The space-time fluid is neither better or worse well behaved than matter.  It is a full description of the Ricci curvature even where $\bar{u}^\alpha$ is not defined.  The Riemann curvature consists only of the Ricci curvature and the Weyl curvature.  The latter is a property of space-time not encapsulated by the space-time fluid.  Weyl curvature interacts with the space-time fluid but also has its own independent dynamics, gravitational radiation being one aspect.  In most of what follows Weyl curvature is left as an external dynamical system.  In many applications to cosmology it is unimportant.

At this stage some readers may think there is nothing new added by the space-time fluid representation of geometrodynamics and be skeptical that anything interesting could come out of it, even pedagogically.  Let's give it a test drive.

\section{Fluid Dynamics: Classical vs.~Space-Time}

This section gives the mathematical representation of fluid mechanics for classical fluids\footnote{
I use "classical fluid" and not "Newtonian fluid" because the latter has a very particular meaning. Newtonian would be more descriptive.} alongside space-time fluids.  The former is likely very familiar to the reader and the latter is a partial representation of geometrodynamics which may be less familiar.  The reason to show both is that the two are nearly identical and not just in a superficial way.\footnote{The close similarity between classical and general relativistic fluids has been noted previously, e.g. in \cite{1990MNRAS.243..509E}.}  It will become tedious how similar they are. This is the point!  Many important phenomena in geometrodynamics can be understood in the language and mathematics of Newtonian physics.

What follows is a sequence of pairs of foundational equations describing classical and space-time fluid dynamics.  Classical equations followed by space-time equations. Admittedly there is a lot of uncertain geometry hidden in the space-time equations (covariant derivatives, etc.) but bear with me. The order of equations forms roughly a derivation. One needn't follow the derivations but note the similarity of the equation pairs.

The 4-velocity of the space-time fluid is $\bar{u}^\alpha$ while for the classical fluid I use the 3-vector $\vec{v}$.  These {\it flows} have {\it streamlines} which are the trajectories which follow flows.  One says {\it fluid elements} move along streamlines.  This does not mean that any or all particles in the fluid necessarily follow this trajectory. A space-time fluid may not even involve particles!  In the fluid description the space-time nevertheless consists of fluid elements which move.  While it is not forbidden that streamlines begin or end in regions of space-time where $\bar{u}^\alpha$ is not defined this is not usually the case. Usually normal matter does not suddenly appear or disappear.

The (proper) time rate of change of a quantity as measured by an observer moving along a streamline is the {\it convective derivatives} defined in the two cases by
\begin{equation}
\dot{f}\equiv\frac{\partial}{\partial t}f+\vec{v}\cdot\vec{\nabla}f
\qquad
\dot{f}\equiv\bar{u}^\alpha f_{;\alpha} .
\end{equation}
Henceforth I reserve the overdot to refer to convective derivatives.  This and the equations that follows are partial differential equations (PDEs). With Lagrangian spatial coordinates (comoving with the fluid) $\dot{f}$ is the partial derivative wrt the time coordinate.  The near identical mathematical description of classical and space-time fluids relates to the PDEs as formulated in Lagrangian coordinates.  This is not to say that space-time fluid is a coordinate (gauge) choice.  This fluid description exists no matter the coordinates used.

Canonical algebraic decomposition of velocity gradients are
\begin{align}
\vec{\nabla}\,\vec{v}&=\frac{1}{3}\,\theta\,\overleftrightarrow{g}
+\overleftrightarrow{\sigma}+\overleftrightarrow{\omega} \qquad
\dot{\vec{v}}=-\vec{\nabla}\Phi+\vec{a}_\mathrm{ng}
\nonumber \\
\bar{u}_{\alpha;\beta}
    &=\frac{1}{3}\,\bar{\cal P}_{\alpha\beta}\,\bar{\theta}
    +\bar{\sigma}_{\alpha\beta}+\bar{\omega}_{\alpha\beta}
    -\dot{\bar{u}}_\alpha\,\bar{u}_\beta
\label{VelocityGradientDecomposition}
\end{align}
where $\Phi$ is the Newtonian gravitational potential and
\begin{align}
\overleftrightarrow{g}&
&\bar{\cal P}_{\alpha\beta}
&\equiv\bar{u}_\alpha\,\bar{u}_\beta+g_{\alpha\beta}  
\nonumber\\
\theta&\equiv\vec{\nabla}\cdot\vec{v}
&\bar{\theta}&\equiv\bar{u}^\alpha_{;\alpha}
\nonumber\\
\overleftrightarrow{\sigma}&\equiv
\frac{1}{2}\,\left(
   \vec{\nabla}\,\vec{v}+(\vec{\nabla}\,\vec{v})^\mathrm{T}
   -\frac{2}{3}\,\theta\,\overleftrightarrow{I}\right)
&\bar{\sigma}_{\alpha\beta}
   &\equiv\frac{1}{2}\,
   {{\bar{\cal P}}_\alpha}^\gamma\left(
   \bar{u}_{\gamma;\delta}+\bar{u}_{\delta;\gamma}
   -\frac{2}{3}\,g_{\gamma\delta}\,{\bar{u}^\epsilon}_{;\epsilon}\right)
   {\bar{{\cal P}}^\delta}_\beta
\nonumber\\
\overleftrightarrow{\omega}&\equiv
   \frac{1}{2}\,\left(
   (\vec{\nabla}\,\vec{v})^\mathrm{T}-(\vec{\nabla}\,\vec{v})\right)
&\bar{\omega}_{\alpha\beta}
    &\equiv\frac{1}{2}\,
   {{\bar{\cal P}}_\alpha}^\gamma\left(
   \bar{u}_{\gamma;\delta}-\bar{u}_{\delta;\gamma}\right)
   \bar{{\cal P}^\delta}_\beta
\nonumber\\
&\vec{a}_\mathrm{ng}=\dot{\vec{v}}+\vec{\nabla}\Phi
&\dot{\bar{u}}_\alpha
&\equiv \bar{u}^\beta\,{\bar{u}^\alpha}_{;\beta}
\label{VelocityGradientComponents}
\end{align}
are respectively the spatial metric, the rates of expansion, shear and rotation and the proper acceleration.  The term "spatial metric" is misused for $\mathcal{P}_{\alpha\beta}$ which is really the spatial projection tensor as is "proper acceleration" for $\vec{a}_\mathrm{ng}$ which is really the non-gravitational acceleration in Newtonian physics.  All 4-vectors and 4-tensors are purely spatial, meaning zero when contracted with $\bar{u}^\alpha$. A similar algebraic decomposition of the stress tensor / stress-energy tensor is
\begin{equation}
\overleftrightarrow{p}=\frac{1}{3}\,p\overleftrightarrow{I}
+\overleftrightarrow{q}
\qquad
T^{\alpha\beta}=\bar{\rho}\,\bar{u}^\alpha\,\bar{u}^\beta
+\bar{p}\,\bar{\mathcal{P}}^{\alpha\beta}
+\bar{q}^{\alpha\beta}
\end{equation}
\label{StressEnergyDecomposition}
where
\begin{align}
p&\equiv\frac{1}{3}\,\mathrm{tr}[\overleftrightarrow{p}]
&\bar{p}&\equiv\frac{1}{3}\,\bar{\mathcal{P}}^{\alpha\beta}\,T_{\alpha\beta}
&\text{pressure}
\nonumber\\
\overleftrightarrow{q}&\equiv\overleftrightarrow{p}-p\,\overleftrightarrow{g}
&\bar{q}^{\alpha\beta}
&\equiv T^{\alpha\beta}-p\,\bar{\mathcal{P}}^{\alpha\beta}
&\text{anisotropic stress tensor}
\label{StressEnergyComponents}
\end{align}
The equations of motion of the fluid are
\begin{align}
\dot{\rho}&=-\rho\,\theta
&
\vec{a}_\mathrm{ng}
&=-\frac{\vec{\nabla}p+\vec{\nabla}\cdot{\overleftrightarrow{q}}}{\rho}
\nonumber \\
\dot{\bar{\rho}}&=-(\bar{\rho}+\bar{p}/c^2)\,\bar{\theta}
    +\bar{q}^{\alpha\beta}\,\bar{\sigma}_{\alpha\beta}/c^2
&
\dot{\bar{u}}^\alpha&=-\frac{{\bar{\cal P}^{\alpha\beta}\,
(\bar{p}_{;\beta}+{{\bar{q}_\beta}^\gamma}}_{;\gamma})}{\bar{\rho}+\bar{p}/c^2}
\label{EnergyMomentumConservation}
\end{align}
which geometrically derives from the contracted Bianchi identity, ${G^{\alpha\beta}}_{;\beta}=0$, which in matter language is the local conservation of mass/energy/momentum, ${T^{\alpha\beta}}_{;\beta}=0$.  From these local conservation laws one can derive the Raychaudhuri equation
\begin{align}
    \dot{\theta}&=-\frac{1}{3}\,\theta^2
    -(2\,\sigma^2-2\,\omega^2-\vec{\nabla}\cdot\vec{a}_\mathrm{ng})
    -4\pi G\,\rho
\nonumber \\
    \dot{\bar{\theta}}&=-\frac{1}{3}\,\bar{\theta}^2
    -(2\,\bar{\sigma}^2-2\,\bar{\omega}^2-{\dot{\bar{u}}^\alpha}_{;\alpha})
    -4\pi G\,(\bar{\rho}+3\,\bar{p}/c^2)
\label{Raychaudhuri}
\end{align}
where 
$\sigma^2\equiv\frac{1}{2}\,\mathrm{tr}[\overleftrightarrow{\sigma}\cdot\overleftrightarrow{\sigma}]$,
$\omega^2\equiv\frac{1}{2}\,\mathrm{tr}[\overleftrightarrow{\omega}\cdot\overleftrightarrow{\omega}]$,  $\bar{\sigma}^2\equiv\frac{1}{2}\,\sigma^{\alpha\beta}\sigma_{\alpha\beta}$ and
$\bar{\omega}^2\equiv\frac{1}{2}\,\omega^{\alpha\beta}\omega_{\alpha\beta}$.
Note that for the classical fluid $\rho$ is the (rest) mass density while for the space-time fluid $\bar{\rho}$ is the mass-energy density which includes rest mass and all other energy.  There is no $\overleftrightarrow{q}\cdot\overleftrightarrow{\sigma}$ term in the mass conservation equation but there would be such a term in a mass+energy conservation equation. Linear shear viscosity produces an anisotropic stress 
$\overleftrightarrow{q}=2\,\mu\,\overleftrightarrow{\sigma}$
or equivalently
$\bar{q}_{\alpha\beta}=2\,\bar{\mu}\,\bar{\sigma}_{\alpha\beta}$ where usually the prefactor is positive.  This is a motivation for referring to the added term in the $\dot{\bar{\rho}}$ equation as a {\it viscosity term} which may be written
$\bar{q}^{\alpha\beta}\,\bar{\sigma}_{\alpha\beta}
=\bar{\mu}\,\bar{\sigma}^2$.  This viscosity indicates an alignment between the shear and the anisotropic stress. For generality allow $\bar{\mu}$ to be positive, negative or zero.

All of these equations are exact for classical and space-time fluids.  There is remarkable similarity between the two. Notable differences are: a modified inertial mass density: $\rho\rightarrow\bar{\rho}+\bar{p}/c^2$; a modified gravitational mass density: $\rho\rightarrow\bar{\rho}+3\,\bar{p}/c^2$; and the addition of a viscosity term.  All these differences explicitly disappear in the $c\rightarrow\infty$ limit.

Neither of these are closed sets of equations but require additional input.  Both fluids require a \emph{matter model} which is a prescription for determining the pressure, anisotropic stress and viscosity. Another common requirement, one which is quite different between the two systems, are the components of the gravitational field not locally determined the matter. In both cases these include tidal forces which affect the evolution of the shear and vorticity.  For the classical fluid these can be determined by the non-local equation $\Phi=\nabla^{-2}4\pi\,G\,\rho$ with boundary conditions at infinity. For the space-time fluid these are local and encoded in the the Weyl curvature tensor. The Weyl tensor has its own dynamics. As will be shown one can infer much about the evolution of the space-time fluid with little knowledge of the tidal fields.

\section{Generalized Friedman Equations and Kurvature}

Space-time fluid elements uniformly expand or contract with no proper acceleration in homogeneous isotropic Friedmann–Lema\^{i}tre–Robertson–Walker (FLRW) space-times. Fluid elements in general space-times expand and contract but also shear, rotate and accelerate and in an inhomogeneous way. In FLRW space-times one quantifies the expansion/contraction by a scale factor $a$. For general space-times one can define a \emph{local scale factor} for each fluid element by
\begin{equation}
\frac{\dot{a}}{a}=\frac{1}{3}\,\theta 
\qquad
\frac{\dot{\bar{a}}}{\bar{a}}=\frac{1}{3}\,\bar{\theta} \ .
\label{LocalScaleFactor}
\end{equation}
These local scale factors are defined only up to a multiplicative constant for each streamline. In expressions for measurable quantities this constant always cancels out.  With this notation the Raychaudhuri equation may be written as a \emph{generalized Friedman equation}
\begin{align}
    \frac{\ddot{a}}{a}
    +\frac{4\pi\,G}{3}\,\rho&=
    -\frac{1}{3}\,
    (\sigma^2-\omega^2-\vec{\nabla}\cdot\vec{a}_\mathrm{ng}) 
\nonumber\\
    \frac{\ddot{\bar{a}}}{\bar{a}}
    +\frac{4\pi\,G}{3}\,
    \left(\bar{\rho}+\frac{3\,\bar{p}}{c^2}\right)&=
    -\frac{1}{3}\,
    (2\,\bar{\sigma}^2-2\,\bar{\omega}^2-{\dot{\bar{u}}^\alpha}_{;\alpha}) \ .
\label{GeneralizedFriedmannEquations}
\end{align}
for each streamline.  Recovering the usual cosmological FLRW Friedman equation requires zero shear, vorticity and proper acceleration.  In the commonly used \emph{separate universe approximation} these terms are assumed negligible in large (super-horizon) volumes but do allow fluid parameters to differ between volumes, i.e.~between separate universes.

The other form of the FLRW Friedman equation is
$\left(\frac{\dot{a}}{a}\right)^2
=\frac{8\pi\,G\,\rho}{3}-\frac{k}{a^2}$
where $k$ is constant and $\frac{k}{a^2}$ is the curvature of the spatial surface of isometry.  This suggests defining a quantity
\begin{equation}
K\equiv\frac{8\pi\,G\,\rho}{3}-\frac{1}{9}\,\theta^2
\qquad
\bar{K}\equiv\frac{8\pi\,G\,\bar{\rho}}{3}-\frac{1}{9}\,\bar{\theta}^2
\label{Kurvature}
\end{equation}
which I refer to as \emph{kurvature} to distinguish it from other quantities spelled curvature.  I will reserve \emph{curvature} to refer to quantities proportional to the kurvature times the scale factor squared, i.e. $\propto a^2\,K$ or $\propto \bar{a}^2\bar{K}$. In FLRW space-times $\bar{K}=\frac{k}{a^2}$ is only a function of time. The more general $K$ and $\bar{K}$ used here can vary both spatially and temporally. Both $K$ and $\bar{K}$ can be considered a measure of the specific gravitational binding energy as is now explained. In a Newtonian context consider a small uniform density ball expanding with a Hubble law: $v=\frac{\dot{a}}{a}\,r$.  One measure of the specific binding energy for matter on the surface of this ball is 
$b=\frac{G\,M}{r}-\frac{1}{2}\,v^2$.  In this case
$K=\frac{1}{2}\,r^2\,b$ where $K>0$ corresponds to bound matter and $K<0$ to unbound matter.  One can apply this binding energy interpretation to every small element of fluid in either the classical or space-time fluid.  When this specific binding energy, $b$, is conserved then so is $a^2\,K$ or $\bar{a}^2\,\bar{K}$. 

$\bar{K}$ is a \emph{local covariant quantity} meaning it is locally measurable. This is true for any property of the space-time fluid which is a measure of the local space-time geometry defined in terms of the local Ricci curvature and its gradients. In contrast $a^2\,K$ and $\bar{a}^2\,\bar{K}$ are not usually local covariant quantities since they depend on the arbitrary multiplicative normalization of the local scale factor. As we shall see one can \emph{sometimes} relate the scale factors to local covariant quantities. 

In "gauge invariant" linear cosmological perturbation theory there is a quantity, usually, denoted $\mathcal{R}$ which is commonly called the \emph{curvature}.  This quantity being dimensionless does not have the dimensions of curvature and it is instead better referred to as the \emph{curvature potential} and denoted here by $\bar{\mathcal{R}}$.  The analogous curvature potential for a classical fluid which I denote $\mathcal{R}$.    As shown in \cite{Stebbins:2026:beta} one can, to linear order, relate $\mathcal{R}$ to $K$ and $\bar{\mathcal{R}}$ to $\bar{K}$ by
\begin{equation}
K=\frac{2}{3}\,\nabla^2\mathcal{R}+\mathcal{O}[2]
\qquad
\bar{K}=\frac{2}{3}\,\nabla^2\bar{\mathcal{R}}+\mathcal{O}[2]
\label{eq:KRrelation}
\end{equation}
where is $\nabla^2$ is "spatial" Laplacian in physical not comoving units. For a classical fluid one can globally define $\nabla^2$ if one chooses a global rest frame while for the space-time fluid specifying $\nabla^2$ requires a choice of time slicing. In any case to linear order one can use 
$\nabla_\mathrm{phys}^2=\frac{1}{{a_\mathrm{FLRW}}^2}\,\nabla_\mathrm{co}^2
+\mathcal{O}[2]$
where $a_\mathrm{FLRW}$ is the background FLRW scale factor and $\nabla_\mathrm{co}^2$ is the Laplacian in terms of comoving coordinates.  Even to linear order and with $\nabla^2$ defined the curvature potential is ambiguous up to addition of solutions of $\nabla^2\mathcal{R}=0$.  Thus in no sense is the curvature potential a locally measurable quantity even for classical fluids.  This is one reason to focus on the kurvature and not on the curvature potential.

\section{Generation of Super-Horizon Curvature Inhomogeneities}
\label{SuperHorizonInhomogeneityGeneration}

One can define a \emph{kurvature density} by
\begin{equation}
    \Delta\rho\equiv\frac{3\,K}{8\pi\,G}
    \qquad
    \Delta\bar{\rho}\equiv\frac{3\,\bar{K}}{8\pi\,G} \ .
\label{KurbatureDensity}
\end{equation}
which is just a different representation of kurvature but has dimensions of density not curvature. $\Delta\bar{\rho}$ gives a natural measure of perturbations from flat FLRW space-times because it is
\begin{itemize}
    \item a non-perturbative quantities
    \item a locally measurable and thus \emph{coordinate independent}
    \item precisely defined for \emph{any} space-time with matter
    \item zero in a flat FLRW space-time.
\end{itemize}
This is not the only such quantity.  Any function of $\bar{\rho}$ and $\bar{\theta}$ satisfy the 1st three criteria, e.g. the \emph{kurvature overdensity}, $\bar{\Delta}\equiv\Delta\bar{\rho}/\bar{\rho}$ or the \emph{generalized density parameter}\footnote{The quantity $\bar{\Delta}\equiv\Delta\bar{\rho}/\bar{\rho}$ looks like $\delta\rho/\rho$ but is a measure of something quite different.}, $\bar{\Omega}\equiv1/(1-\bar{\Delta})$, which reduces to the normal density parameter in FLRW space-times.  $\bar{\Delta}$ and $\bar{\Omega}-1$ also satisfy the fourth criteria making them good perturbation variables which are also dimensionless. Here I use $\Delta\bar\rho$ which is easier to work with and is relevant since kurvature describes the dominant type of primordial inhomogeneity found in our universe. Perturbatively eq.~\ref{eq:KRrelation} gives Poisson-like relations $c^2\,\nabla_\mathrm{co}^2\,\mathcal{R}=4\pi\,{a_\mathrm{FLRW}}^2\,G\,\Delta\rho$
and
$c^2\,\nabla_\mathrm{co}^2\,\bar{\mathcal{R}}=4\pi\
{a_\mathrm{FLRW}}^2\,G\,\Delta\bar{\rho}$.

The evolution of kurvature density along streamlines, given by 
eq.s~\ref{EnergyMomentumConservation}~\&~\ref{Raychaudhuri}, is
\begin{align}
\dot{\Delta\rho}+2\,\frac{\dot{a}}{a}\,\Delta\rho&=
\frac{\dot{a}}{a}\,
\frac{2\,\sigma^2-2\,\omega^2-\vec{\nabla}\cdot \vec{a}_\mathrm{ng}}{4\pi\,G}
\nonumber \\
\dot{\Delta\bar{\rho}}+2\,\frac{\dot{\bar{a}}}{\bar{a}}\,\Delta\bar{\rho}&=
\frac{\dot{\bar{a}}}{\bar{a}}\,
\frac{2\,\bar{\sigma}^2-2\,\bar{\omega}^2-{\dot{\bar{u}}^\alpha}_{;\alpha}}
{4\pi\,G}
-\frac{\bar{\mu}\,\bar{\sigma}^2}{c^2} \ .
\label{KurvatureDensityEquationOfMotion}
\end{align}
One recovers the FLRW solution $\bar{K}\propto\Delta\bar{\rho}\propto a^{-2}$ when the right hand side is zero, (no shear, vorticity, proper acceleration, viscosity) so the specific binding energy, $b$, is constant.


Henceforth refine the question of \S\ref{InhomogeneityGenerationQ} to "\emph{Can curvature inhomogeneities, $\Delta\bar{\rho}$ , be generated on super-horizon scales?} or more specifically does eq.~\ref{KurvatureDensityEquationOfMotion} allow for $\Delta\bar{\rho}$ to be generated spontaneously on super-horizon scales?

Formal exact solutions of eq.~\ref{KurvatureDensityEquationOfMotion} are given by line-integrals along stream-lines
\begin{align}
\Delta\rho_\mathrm{f}&=
\frac{{a_\mathrm{i}}^2}{a_\mathrm{f}^2}\,\Delta\rho_\mathrm{i}+
\int_{t_\mathrm{i}}^{t_\mathrm{f}}dt\,
\frac{\dot{a}\,a}{a_\mathrm{f}^2}\,
\frac{2\,\sigma^2-2\,\omega^2-\vec{\nabla}\cdot\vec{a}_\mathrm{ng}}{4\pi\,G}
\nonumber \\
\Delta\bar{\rho}_\mathrm{f}&=
\frac{{\bar{a}_\mathrm{i}}^2}{{\bar{a}_\mathrm{f}}^2}\,
\Delta\bar{\rho}_\mathrm{i}+
\int_{\tau_\mathrm{i}}^{\tau_\mathrm{f}}d\tau\,
\frac{\dot{\bar{a}}\,\bar{a}}{{\bar{a}_\mathrm{f}}^2}\,
\frac{2\,\bar{\sigma}^2-2\,\bar{\omega}^2-{\dot{\bar{u}}^\alpha}_{;\alpha}}
{4\pi\,G}
-
\int_{\tau_\mathrm{i}}^{\tau_\mathrm{f}}d\tau\,
\frac{{\bar{a}}^2}{{\bar{a}_\mathrm{f}}^2}\,
\frac{\bar{\mu}\,\bar{\sigma}^2}{c^2}
\label{LineIntegralSolution}
\end{align}
where "f" is final, "i' initial and $\tau$ is the proper time along the streamline. The undetermined multiplicative constants in $a$ and $\bar{a}$ cancel out.  
While the integrals of eq.~\ref{LineIntegralSolution} are generally non-zero in many contexts they are negligible so $\mathcal{R}\propto a^2\,\Delta\rho$ or  $\bar{\mathcal{R}}\propto \bar{a}^2\,\Delta\bar{\rho}$ are approximately constant so this approximation is \emph{constant curvature}.  If the integrals are significantly non-zero then \emph{curvature is generated}.  In the separate universe approximation all the terms in the integrand are neglected so this is a constant curvature approximation.  Note that, unlike $K$ or $\bar{K}$ the curvature quantities $a^2\,K$ or $\bar{a}^2\bar{K}$ are not a measurable quantities because of the undefined multiplicative factors in the scale factor. One cannot even compare their value from one streamline to another since the scale factors can be normalized differently along different streamlines.  This ambiguity encompasses the time-slicing coordinate (gauge) ambiguity of cosmological perturbation theory.  

For the space-time fluid if there is a conserved current 
$(\bar{s}\,\bar{u}^\alpha)_{;\alpha}=0$ where $\bar{s}$ is a locally measurable quantity then $\bar{s}\propto\frac{1}{\bar{a}^3}$ and one can always define the scale factor by $\bar{a}\equiv(s_\mathrm{fid}/\bar{s})^{1/3}$ for any fiducial density $s_\mathrm{fid}$.  In this case $\frac{\Delta\bar{\rho}}{\bar{s}^{2/3}}$ is locally measurable and a constant of motion in constant curvature scenarios. In the case of a barotropic (fixed $\bar{p}[\bar{\rho}]$) perfect ($\bar{q}^{\alpha\beta}=0$) space-time fluid there is such a current given by
\begin{equation}
\bar{s}\equiv\ln\left[\int^{\bar{\rho}}_{\rho_\mathrm{fid}}
\frac{d\rho}{\rho+p[\rho]\,c^2}\right]\ .
\end{equation}
which is the fluid entropy density\footnote{The conserved fluid entropy density is related to the gauge invariant linear theory $\zeta=\frac{1}{3}\,\frac{\delta\rho}{\rho+p}$ \cite{PhysRevD.68.103515}.} for any (universal) choice of fiducial density $\rho_\mathrm{fid}$.  Since for a classical fluid $c\rightarrow0$ so $s\propto\rho$ and the fluid entropy reduces to conserved (rest) mass density and the measurable constant of motion in constant curvature scenarios is $\frac{\Delta\rho}{\rho^{2/3}}$.

One can use the divergence theorem to "integrate out" $\vec{\nabla}\cdot\vec{v}$ or ${\dot{\bar{u}}^\alpha}_{;\alpha}$ from volume integrals of eq.~\ref{KurvatureDensityEquationOfMotion}.  The divergence of the proper acceleration in classical and space-time fluids differ in that the former is 3D spatial divergence and the latter is a 3+1-D space-time divergence. For classical fluids the divergence theorem involves a 3D volume with a 2D boundary and for a space-time fluid a 4D volume with a 3D boundary. To make the volume integrals of the two types of fluid comparable one can choose a spatial hypersurface (time-slice) in space-time to compare with a constant time surface in the classical fluid and then integrate over an infinitesimally thin 4D volume surrounding this surface. The results will depend somewhat on the choice of time-slice.  If the spacetime fluid is irrotational, i.e. $\bar{\omega}=0$ everywhere, then there is a unique foliation of space-time I call \emph{flow orthogonal time slicing} (FOTS) where the leaves are spatial hypersurfaces whose normal is $\bar{u}^\alpha$. In this case one can choose the 4D volume to be a thin slice of proper time thickness $\delta\tau$ in the $\bar{u}^\alpha$ frame.  Since $\bar{u}^\alpha\dot{\bar{u}}_\alpha=0$ the past and future boundary terms for this volume are zero in the limit $\delta\tau\rightarrow0$.  For both fluids only the spatial boundary contributes.  If the spatial sections are compact there is no spatial boundary.  Alternately, if $\dot{\bar{u}}^\alpha\ne0$ is confined to a finite spatial region and one takes the spatial boundary to encompass this volume then the spatial boundary term is zero.  Alternately still, for large enough volumes, unless there is a systematic inward or outward proper acceleration at large distances, the 2D surface integral will be much smaller than the 3D volume integral and may be neglected.  In either case if one ignores the spatial boundary one finds\footnote{
Since pairs of leaves of FOTS have spatially varying separation in $\tau$ one cannot simply use $\delta\tau\rightarrow\int d\tau$ to integrate over a finite 4D volume.}
\begin{equation}
\int d^3V\,\left(
    \frac{\dot{(a^2\,\Delta\rho})}{a\,\dot{a}}
    -\frac{\sigma^2}{2\pi\,G}\right)=0
\qquad
\delta\tau \int d^3\bar{V}\,\left(
    \frac{\dot{(\overline{a}^2\,\Delta\bar{\rho}})}{\bar{a}\,\dot{\bar{a}}}
    -\frac{\bar{\sigma}^2}{2\pi\,G}
    +\frac{\bar{a}}{\dot{\bar{a}}}\,\frac{\,\bar{\mu}\,\bar{\sigma}^2}{c^2}
    \right)=0 \ .
\label{AverageCurvatureEvolutionIrrotational}
\end{equation}
where I have taken $\omega=\bar{\omega}=0$ and used the 4D volume element 
$\sqrt{-g}\,d^4x^\alpha=d\tau\,d^3\bar{V}$ where $d^3\bar{V}$ is the spatial volume element in the $\bar{u}^\alpha$ frame.  The reason $\bar{\omega}=0$ has been assumed is that it provides a convenient (FOTS) choice of spatial hypersurface for a space-time fluid.  For direct comparison with a classical fluid I have also taken $\omega=0$.  The main value added by eq.~\ref{AverageCurvatureEvolutionIrrotational} over eq.~\ref{LineIntegralSolution} is that  $\vec{\nabla}\cdot\vec{a}_\mathrm{ng}$ and ${\dot{\bar{u}}^\alpha}_{;\alpha}$ have been rigorously eliminated by spatial averaging.

The terms which contribute to curvature generation: shear, vorticity and viscosity, are generally either positive or negative definite and cannot be averaged to zero.  In the case of the space-time fluid there is the possibility of cancellation of shear, vorticity and viscosity.  For perfect fluids the viscosity is zero and if vorticity is zero initially then it remains zero, i.e. $\omega=0$ or $\bar{\omega}=0$.  Perfect fluids provide a physical model where one can completely eliminate rotation and viscosity curvature generation is $\propto\theta\,\sigma^2$ or $\propto\bar{\theta}\,\bar{\sigma}^2$. Since $\sigma^2,\,\bar{\sigma}^2\ge0$ for a perfect fluid, wherever there is non-zero shear, positive curvature will be generated in expanding fluid fluid elements and negative shear in contracting fluid elements. At early times our universe contains something close to a perfect fluid and is expanding nearly everywhere.  Any inhomogeneity no matter how small will result in non-zero shear and thus systematically generate of positive curvature, i.e. the matter will become more bound.

N.B. there is no conservation of curvature in the sense of a conserved current even where curvature is constant. Thus there is no conserved global curvature charge in analog with a conserved energy.  In this sense constant curvature is not an analog of conservation of energy. If one reformulates the question of \S\ref{InhomogeneityGenerationQ} as \emph{Do space-time fluid elements have constant curvature in some average sense?} the answer is {\bf NO}.  Curvature is constantly being generated on all large scales.

\section{Answers and Questions and Answers}

\subsection{Answers}

This essay has so far addressed two nagging questions.  The 1st is whether there is a sensible way to formalize the idea that space moves.  The answer proposed here is rather trivial: that space inherits velocity from the matter it contains and more generally suggests that Ricci curvature be described as a fluid not to be distinguished from matter.  If, as Einstein's equations motivates, one considers matter stress-energy and curvature as the same thing these are rather reasonable descriptive terms.  I have found the space-time fluid description a good conceptual framework for describing some aspects of geometrodynamics.  

The 2nd question is whether one can generate inhomogeneities on super-horizon scales.  This is relevant to the assertion that observed large scale cosmic inhomogeneities prove that well separated parts of the universe were in causal contact in the past.  I address this question using space-time fluid language.  It turns out this is not so much a question of energy conservation but rather whether or not net curvature can be generated on large scales (curvature is a measure of specific gravitational binding energy).  I conclude that generation of curvature (non-constant curvature) can be important for generating large scale inhomogeneities outside the horizon.

I also show that, while space-time fluid dynamics does not describe all of geometrodynamics, where it does, it is extremely similar to classical fluid dynamics.  Curvature can be defined as a measure of specific gravitational binding energy in exactly the same way for classical and space-time fluids. Both exhibit nearly identical generation of curvature. Therefore generation of curvature is a purely classical phenomenon. In fact it is argued below that it is not fundamentally a gravitational phenomenon either.

In the course of this investigation many other questions arose.  Some of those which I have resolved are addressed in what follows.

\subsection{Cosmic Anarchy and Kurvature}
\label{Anarchy}

\noindent Use \emph{cosmic anarchy} to refer to the following concepts
\begin{itemize}
    \item \emph{flow anarchy:} The equations of classical fluid mechanics do not pose any restrictions on the fluid velocity field.  For any flow in Euclidean space there will be densities and pressures and anisotropic stresses which are consistent with this flow and the equations of fluid mechanics.
    \item \emph{geometric anarchy:} Einstein's equations do not pose any restrictions on space-time geometry. For any space-time geometry there will be densities and pressures and anisotropic stresses which are consistent with any geometry and Einstein's equation.
\end{itemize}
What allows fluid mechanics and GR to be predictive, say in the sense of allowing there to be well-posed initial value problems, is the restriction to "realistic" matter models e.g.~a realistic relation of pressure to density or a realistic viscosity model for anisotropic stresses.  Only with these restrictions does one restrict fluid flows or space-time geometry.

\vskip10pt

\noindent{\bf Questions:}

\begin{itemize}

\item[\bf Q1]
    Is there \emph{kurvature anarchy}?  If one starts with some initial conditions are there any constraints on the final condition of the fluid? Or is anything possible?

\begin{itemize}
        \item[\bf A1a] For a classical fluid the curvature is $K=\frac{8\pi\,G\,\rho}{3}-\frac{1}{9}\,\theta^2$. $\theta$ being the divergence of $\vec{v}$ is anarchic as part of flow anarchy. So the question becomes: For fixed initial $\rho_\mathrm{i}$ and $\vec{v}_\mathrm{i}$ and fixed final $\vec{v}_\mathrm{f}$ are there any constraints on the final $\rho_\mathrm{f}$?  By counting the much smaller number of degrees of freedom in $\rho_\mathrm{f}$ relative to the infinitely greater number of degrees of freedom in $\vec{v}$ for $t_\mathrm{i}<t<t_\mathrm{i}$ anything should be possible, but with a caveat.
        \begin{itemize}
            \item Since mass is conserved one requires the total mass not be changed.
            \begin{itemize}
                \item Mass conservation may not be an issue if the total mass is infinite.
            \end{itemize}
        \end{itemize}

     \item[\bf A1b] Is space-time kurvature anarchic as part of geometric anarchy.

    \item[\bf A1c] Kurvature is anarchic because kurvature is defined by the Ricci curvature which is a component of the anarchic space-time geometry.

    \end{itemize}
\end{itemize}

\subsection{Secular Trends}
\label{Secular}

While anarchy means that "anything is possible" it is nevertheless true that various hydrodynamical/geometrical properties are correlated with each other when these properties are not really distinct entities.  These correlations lead to secular trends in, for example, the generation of curvature depending on whether the fluid is expanding or contracting: 

\begin{itemize}
\item[\bf Q2]
Is generated curvature preferentially positive or negative?

    \begin{itemize}
        
        \item[\bf A2a] Gravity is time reversal invariant so there is no preference is this sense.
        
        \item[\bf A2b] However there can and will be secular trends for most flows of interest.  The contributions to curvature generation from eq.~\ref{KurvatureDensityEquationOfMotion} or \ref{LineIntegralSolution} are
        \begin{align*}
            &                           & \text{shear}  & &\text{vorticity} 
                                                        & & \text{viscosity} 
                                                        & & \text{force}\\
            &\text{classical fluid}     & \propto+\theta\,\sigma^2 &
                                        & \propto-\theta\,\omega^2 & &0&
                        & \propto-\theta\,\vec{\nabla}\cdot\vec{a}_\mathrm{ng}\\
            &\text{space-time fluid}    & \propto+\bar{\theta}\,\bar{\sigma}^2 &
                                        & \propto-\bar{\theta}\,\bar{\omega}^2 &
                                        & \propto-\bar{\mu}\,\bar{\sigma}^2    &
                                        & \propto
                            -\bar{\theta}\,{\dot{\bar{u}}^\alpha}_{;\alpha}
        \end{align*}
        The force contribution can be can be averaged to zero as in eq.~\ref{AverageCurvatureEvolutionIrrotational}.  The contribution of shear and vorticity are positive and negative for expanding fluids and negative and positive for contracting fluids.  The viscosity contribution depends on the sign of the coefficient $\bar{\mu}$. One would expect this viscosity coefficient to be positive corresponding to alignment between shear and anisotropic stress but anti-alignment is not impossible.  While there is no obvious relation between viscosity and the other terms there is between shear and vorticity: 

        \begin{itemize}
          
            \item Any flow can be decomposed into irrotational ("p" for potential) flow and rotational ("r") flow, the former having zero vorticity and the latter zero expansion.  In Euclidean geometry identities relate spatial averages\footnote{This assumes there is no net shear or rotation at infinity.}
            $\overline{\sigma^2}-\overline{\omega^2}=\overline{\sigma_\mathrm{p}^2}$ 
            where $\sigma_\mathrm{p}$ is the shear from irrotational (potential) flow. Thus on average the rotational flow does not contribute to generation of curvature for classical fluids.  More generally spatial averages of curvature generation are given by
            \begin{equation}
                \overline{\ \frac{\dot{(a^2\,\Delta\rho})}{a^2\,\theta}\ }
                =\frac{\overline{\sigma_\mathrm{p}^2}}{6\pi\,G} \ .
            \label{AverageCurvatureEvolutionClassical}
            \end{equation}
            For classical fluids an average generated curvature is positive for an expanding fluid and negative for a contracting fluid.
            
            \item For a space-time fluid\footnote{I'm being fuzzy about which time-slice the space-time fluid average is over.}
            \begin{equation}
                \overline{\ \frac{\dot{(\overline{a}^2\,\Delta\bar{\rho}})}
                {\bar{a}^2\,\bar{\theta}}\ }
                =\frac{
                \overline{\,\bar{\sigma}_\mathrm{p}^2\,}+
                \overline{\,
                    \bar{\sigma}_\mathrm{r}^2-\bar{\omega}_\mathrm{r}^2\,}}
                {6\pi\,G}
                -\overline{\ \frac{\bar{\mu}\,
                (\bar{\sigma}_\mathrm{p}^2+\bar{\sigma}_\mathrm{r}^2)}
                {\bar{\theta}\,c^2}\ }  \ .
            \label{AverageCurvatureEvolutionSpacetime}
            \end{equation}
            which may be positive or negative.
            \item In cosmological applications the FOTS spatial sections are close to Euclidean so one expects
            $\overline{\,\bar{\sigma}_\mathrm{p}^2\,}\gg
            \overline{\,\bar{\sigma}_\mathrm{r}^2-\bar{\omega}_\mathrm{r}^2\,}$, i.e.~there is approximate but not be exact cancellation between 
            $\overline{\,\bar{\sigma}_\mathrm{r}^2\,}$ and 
            $\overline{\,\bar{\omega}_\mathrm{r}^2\,}$ as there is in flat space.\footnote{This is not always the case, e.g.~in the Godel space-time $\theta=\sigma=0$ but $\omega\ne0$.} Thus the sign of the generated curvature is a competition between irrotational shear and viscosity which has the opposite signs when $\bar{\mu}>0$.  For an expanding universe the shear increases curvature and the viscosity decreases it.  If the anisotropic stress is small then so is the viscosity and the average curvature will increase.   
            
            \item As an example consider a classical fluid with flat FLRW initial conditions: expanding homogeneous density $\rho_\mathrm{i}$ and $\Delta\rho_\mathrm{i}=0$.  If one restricts the flow to have a minimum time dependent expansion rate everywhere $\theta>\theta_\mathrm{min}[t]>0$ and non-zero rms shear $\overline{\sigma_\mathrm{p}^2}>\sigma_\mathrm{min}^2[t]\ge0$ then one can show that
            \begin{equation}
                \overline{\,\Delta\rho\,}\ge
                \frac{1}{6\pi\,G}
                \left(\frac{\rho}{\rho_\mathrm{i}}\right)^{2/3}
                \int_{t_\mathrm{i}}^t dt'\,
                \sigma_\mathrm{min}[t']^2\,\theta_\mathrm{min}[t']\,
                e^{\frac{2}{3}\int_{t_\mathrm{i}}^{t'} dt''\,
                    \theta_\mathrm{min}[t'']} \ .
            \end{equation}
            and similarly for a contracting fluid with 
            $\theta<\theta_\mathrm{max}[t]<0$
            \begin{equation}
                \overline{\,\Delta\rho\,}\le
                \frac{-1}{6\pi\,G}
                \left(\frac{\rho}{\rho_\mathrm{i}}\right)^{2/3}
                \int_{t_\mathrm{i}}^t dt'\,
            \sigma_\mathrm{min}[t']^2\,(-\theta_\mathrm{max}[t'])\,
                e^{\frac{2}{3}\int_{t_\mathrm{i}}^{t'} dt''\,
                    \theta_\mathrm{min}[t'']} \ .
            \end{equation}
            By taking the limits $\theta_\mathrm{min}[t]\rightarrow0^+$ or $\theta_\mathrm{max}[t]\rightarrow0^-$ one can make the general statement that the generated curvature is positive for expanding fluids and negative for contracting fluids. 
            
            \item One should be able to make a similar arguments for the space-time fluid when viscosity is sub-dominant.

     \end{itemize} 

        \item[\bf A2c] Consider a spatial region $\mathfrak{R}$ on a classical constant $t$ surface or on a FOTS time slice for space-time fluid assuming $\bar{\omega}=0$.   This is what was as used in eq.~\ref{AverageCurvatureEvolutionIrrotational}.  Define spatial averages over the region $\mathfrak{R}$:
        \begin{eqnarray}
            \overline{\rho_\mathfrak{R}}&\equiv&
            \frac{\int_\mathfrak{R} d^3V\,\rho}{\int_\mathfrak{R} d^3V} \qquad  
            \overline{\theta_\mathfrak{R}}\equiv
            \frac{\int_\mathfrak{R} d^3V\,\theta}{\int_\mathfrak{R} d^3V} \qquad
            \overline{\theta^2_\mathfrak{R}}\equiv
            \frac{\int_\mathfrak{R} d^3V\,\theta^2}{\int_\mathfrak{R} d^3V}
            \nonumber\\
            \overline{\bar{\rho}_\mathfrak{R}}&\equiv&
            \frac{\int_\mathfrak{R} d^3\bar{V}\,\bar{\rho}}
            {\int_\mathfrak{R} d^3\bar{V}} \qquad  
            \overline{\bar{\theta}_\mathfrak{R}}\equiv
            \frac{\int_\mathfrak{R} d^3\bar{V}\,\bar{\theta}}
            {\int_\mathfrak{R} d^3\bar{V}} \qquad
            \overline{\bar{\theta}^2_\mathfrak{R}}\equiv
            \frac{\int_\mathfrak{R} d^3\bar{V}\,\bar{\theta}^2}
            {\int_\mathfrak{R} d^3\bar{V}}
        \label{RegionalAverages}
        \end{eqnarray}
        so that the volume averaged kurvature densities are
        \begin{align}
            \overline{\Delta\rho_\mathfrak{R}}&\equiv
            \frac{\int_\mathfrak{R} d^3V\,\Delta\rho}{\int_\mathfrak{R} d^3V}
            =\overline{\rho_\mathfrak{R}}
                        -\frac{\overline{\theta^2_\mathfrak{R}}}{24\pi\,G}
            =\Delta\rho_\mathfrak{R}
    -\frac{\overline{\theta^2_\mathfrak{R}}-\overline{\theta_\mathfrak{R}}^2}
            {24\pi\,G}
            &
            \Delta\rho_\mathfrak{R}&\equiv\overline{\rho_\mathfrak{R}}
                        -\frac{\overline{\theta_\mathfrak{R}}^2}{24\pi\,G}
            \nonumber\\
             \overline{\Delta\bar{\rho}_\mathfrak{R}}&\equiv
            \frac{\int_\mathfrak{R} d^3\bar{V}\,\Delta\bar{\rho}}
                 {\int_\mathfrak{R} d^3\bar{V}}
            =\overline{\bar{\rho}_\mathfrak{R}}
                        -\frac{\overline{\bar{\theta}^2_\mathfrak{R}}}{24\pi\,G}
            =\Delta\bar{\rho}_\mathfrak{R}
            -\frac{\overline{\bar{\theta}^2_\mathfrak{R}}-
                   \overline{\bar{\theta}_\mathfrak{R}}^2}
            {24\pi\,G}
            &
            \Delta\bar{\rho}_\mathfrak{R}&\equiv\overline{\rho_\mathfrak{R}}
                        -\frac{\overline{\bar{\theta}_\mathfrak{R}}^2}{24\pi\,G}
        \label{RegionalCurvatureDensity}
        \end{align}
        where $\Delta\bar{\rho}_\mathfrak{R}$ and $\Delta\rho_\mathfrak{R}$ 
        are \emph{coarse grained kurvature densities} as determined from the volume averaged density and expansion. If there is any inhomogeneity in $\theta$ or in $\bar{\theta}$ then $\Delta\rho_\mathfrak{R}>\overline{\Delta\rho_\mathfrak{R}}$ or $\Delta\bar{\rho}_\mathfrak{R}>\overline{\Delta\bar{\rho}_\mathfrak{R}}$.
        Thus \emph{the coarse grained kurvature is greater than the volume averaged kurvature}.
        \item Eq.s~\ref{RegionalAverages} \& \ref{RegionalCurvatureDensity} for classical and space-time fluid equations are identical but there are important differences between $\overline{\rho_\mathfrak{R}}$ and $\overline{\Delta\bar{\rho}_\mathfrak{R}}$ and between $\overline{\theta_\mathfrak{R}}$ and $\overline{\bar{\theta}_\mathfrak{R}}$.
        \begin{itemize}
            \item
            Mass is conserved for classical fluids while the space-time fluid mass-energy is not.\footnote{An exception occurs for barotropic perfect fluids when $\bar{s}\propto\bar{\rho}$ which generally only occurs for dust $\bar{p}=0$.}  
            \item
            The mass and mass-energy  within $\mathfrak{R}$ for the fluids are given by
            $M_\mathfrak{R}=\overline{\rho_\mathfrak{R}}\,V_\mathfrak{R}$ or
            $\bar{M}_\mathfrak{R}=\overline{\bar{\rho}_\mathfrak{R}}\,
            \bar{V}_\mathfrak{R}$ where 
            $V_\mathfrak{R}\equiv\int_\mathfrak{R} d^3V$ or
            $\bar{V}_\mathfrak{R}\equiv\int_\mathfrak{R} d^3\bar{V}$
            is the volume of $\mathfrak{R}$.  In classical flat space one can determined $V_\mathfrak{R}$ from the boundary, $\partial{V}_\mathfrak{R}$, and perhaps $M_\mathfrak{R}$ from an inventory of mass which entered $\mathfrak{R}$.  Thus for a classical fluid one can determine $\overline{\rho_\mathfrak{R}}$ "from without", by knowing $M_\mathfrak{R}$ and $\partial{V}_\mathfrak{R}$. For a space-time fluids where space is not flat one can neither know  $\bar{V}_\mathfrak{R}\equiv\int_\mathfrak{R} d^3\bar{V}$ from $\partial\bar{V}_\mathrm{R}$ or know the mass energy
            $\bar{M}_\mathfrak{R}\equiv\int_\mathfrak{R} d^3\bar{V}\,\bar{\rho}$ from an inventory of what entered $\mathfrak{R}$.
            \item
            For a classical fluid if one tracks a comoving volume, $\mathfrak{R}$, over time then
            $\overline{\theta_\mathfrak{R}}=\frac{d}{dt}\ln[V_\mathfrak{R}]$ which one can also determine by tracking the change in the boundary of $\mathfrak{R}$.  This gives the change of volume.  Thus "from without" one can determine $\overline{\rho_\mathfrak{R}}$ and $\overline{\theta_\mathfrak{R}}$ which is all that is requires to determine the coarse grained curvature density which places a upper bound on the volume averaged curvature density. Again the curved space of the space-time fluid prevents one from making these inferences.
            \item
            An idealized \emph{isolated compensated cosmological inhomogeneity} is one where outside of some region, call it $\mathfrak{R}$, the flow is exactly FLRW.  For classical fluids this requires that there be no excess or deficit of mass in $\mathfrak{R}$ relative to the surrounding FLRW.  If the background is flat then the coarse-grained curvature density is $\Delta\rho_\mathfrak{R}=0$ so that
            $\overline{\Delta\rho_\mathfrak{R}}<0$ if there is any inhomogeneity in $\theta$. This statement that for compensated inhomogeneities the volume averaged kurvature is negative is not surprising given that more of the volume is, by definition, in underdense regions which likely have negative binding energy.  It is interesting you can show this in complete generality.
            \item
            I conjecture these properties of classical fluids apply also to space-time fluids in many cases of interest.
        \end{itemize}
    \end{itemize} 

\end{itemize}

\subsection{Reality Check}

\begin{itemize}
    \item[{\bf Q3}] If the curvature is constantly increasing does that mean that a flat universe evolves into a closed universe?  Certainly that is topologically impossible!
    \begin{itemize}
        \item[{\bf A3}] It is indeed impossible.  For a spatially isotropic homogeneous space-time, $\overline{\,\Delta\bar{\rho}\,}>0$, does imply a $\mathbb{S}^3$ spatial geometry or compactification thereof.  This is topologically inconsistent with initial spatially infinite space-times.  However both topologies can have positive, negative or zero $\overline{\,\Delta\bar{\rho}\,}$ for anisotropic or inhomogeneous space-times. The requirement to generate curvature is generally $\bar{\sigma}\ne0$ which means that space-time is locally anisotropic.
    \end{itemize}

    \item[{\bf Q4}] What about the speed of light?  It doesn't play a role in the classical fluid equations as it shouldn't but also doesn't play a role in the space-time fluid equations. Is that correct?
    \begin{itemize}
         \item[{\bf A4}] Yes that is correct.  Apart from the pressure correction to the inertial and gravitational mass density there is no $c$ and thus the space-time fluid dynamics in itself doesn't know about $c$. A sound speed is important but the speed of light is not.\footnote{Matter models often do depend on $c$.}  However the space-time fluid is coupled to the Weyl curvature whose dynamics does know about $c$.
    \end{itemize}
    
    \item[{\bf Q5}] If one takes the space-time fluid model seriously, down to the molecular scale where the shear is probably enormous, won't that lead to enormous generation of curvature?
    \begin{itemize}
         \item[{\bf A5}] I don't think so.  Remember that curvature increases when there is shear and expansion but decreases when there is shear and contraction.  On the molecular scales the expansion / contraction, which is the divergence of the time-like eigenvector of the stress energy tensor, fluctuates about zero. I believe these small scale fluctuations should cancel out such that the molecular scale does not play much of a role in overall increase or decrease of gravitational binding energy.  The same should be true for cosmological matter in the "stable clustering regime" where there is no systematic expansion or contraction.  This is in contrast to larger cosmological scales which are highly skewed toward expansion.  These large scales is where one would expect systematic positive curvature generation.
    \end{itemize}

    \item[{\bf Q6}]
    Are there "curvature equilibrium" space-times where, perhaps in some statistical sense, curvature neither increases or decreases?
    \begin{itemize}
         \item[{\bf A6}]
         Geometric anarchy suggests one could invent such a space-time.  As for space-times with "realistic" matter:
         \begin{itemize}
            \item In cases where gravity is unimportant all of the normal equilibrium states should have no net curvature generation.
            \item Where gravity is important I am not sure.
         \end{itemize}
    \end{itemize}
\end{itemize}

\subsection{Eulerian Thinking}

 \begin{itemize}    
    \item[{\bf Q7}]
    A classical fluid should obey local mass conservation and not produce large scale inhomogeneities according eq.~\ref{rhofromp}.  How can it produce large scale inhomogeneities here?
    \begin{itemize}
        \item[{\bf A7}]
        It is true that there is no large scale density inhomogeneities generated in a classical fluid. What is discussed here is generation of curvature on large scales not of mass.  It is clear from eq.~\ref{RegionalCurvatureDensity} that the two, while related, are not the same.   It is claimed here that curvature is the relevant type of inhomogeneity because this is what is observed in cosmology.
    \end{itemize}
    
    \item[{\bf Q8}]
    If you write down space-time fluid dynamics in Eulerian coordinates will mass generation go away as it does for classical fluids? 
    \begin{itemize}
         \item[{\bf A8}]
         This goes to the heart of the matter! There are no Eulerian coordinates in GR.  Newtonian physics has a rigid background geometry unmoved by the matter.  GR does not have this additional structure, geometry is deformable not rigid, since matter and geometry are part and parcel of the same thing one cannot remove the deformations (of Lagrangian coordinates) without removing the matter.
    \end{itemize}
    
    \item[{\bf Q9}]
    Can't there be conserved quantities in GR? 
    \begin{itemize}
         \item[{\bf A9}]
         Yes there can be and often are, e.g.~entropy or particle number. In most cases these are not directly related to curvature as $\Delta\rho$ is. Furthermore what one often requires from a conserved quantity is constancy of the volume weighted average of that quantity. Unlike in Newtonian physics under geometrodynamics the volume is dynamical and can fluctuate so even a conserved quantity can develop large scale inhomogeneities if there develop large scale inhomogeneities in the volume.
    \end{itemize}
    
    \item[{\bf Q10}]
    Isn't space-time more rigid than the comoving Lagrangian coordinates used to describe the space-time fluid?
    \begin{itemize}
         \item[{\bf A10}]
         The space-time fluid is not a coordinate system!  Its dynamics might be mostly simply expressed in Lagrangian coordinates but one can use whatever coordinates one wants.  In other coordinate systems curvature is not static but moves. Even though the space-time fluid can have rapid and complex motion the gravitational fields are (usually) weak ($\mathrm{length}^2\times G_{\alpha\beta}\ll1$) so geometry differs only slightly from flat space.  The motion of the space-time fluid is more like the motion of the waves on the ocean which may seem rapid and chaotic but are generally insignificant relative to the ocean itself.
         \begin{itemize}
             \item
             Different coordinates will accentuate deviations from flat geometry to different extents.
             \item
             The structure of flat space includes isometries which provides time-like Killing vectors which are not parallel to $\bar{u}^\alpha$ and provide not only conserved energy but also a "rigid" structure of space-time.  Rigid meaning that flows following Killing vectors do not have expansion, shear or vorticity.  Exact FLRW space-times usually don't have time-like Killing vectors but do have conformal Killing vectors which again is a rigid structure of space-time with no shear or vorticity.  If there are such rigid structures by all means use them.  Generally there is not.  Coordinates often used in cosmological perturbation theory approximate this rigidity and are far more rigid than Lagrangian coordinates. 
             \item
             We like (or should like) rigid coordinates because these are how we traditionally speak of physical phenomena and also because we are familiar with an environment where there are rigid structures, such as the Earth.\footnote{If one lived completely on the high seas would one have a different perspective?}  If space-time rigidity is only approximate it may be impossible to precisely describe the universe using the language we are used to using.  For example in describing the large scale structure of the universe one might say that matter flows from voids into filaments, i.e. matter moves from here to there.  Can one precisely define what "here" and "there" is? 
         \end{itemize}
    \end{itemize}
    
\end{itemize}

\section{The Here/There Problem}
\label{HereThere}

It is fine to say that a clump of matter which is \emph{here now} was \emph{there then}.  These are events not places.  An issue arises in general relativity when one tries to specify where \emph{there} is \emph{now}. Propagating spatial positions forward in time can be thought of as a specification of how space moves which is the 1st of the questions addressed in this essay.  Coordinates are such a specification - both \emph{here} and \emph{there} keep the same spatial coordinate value.  However coordinates are mostly a convenience.  Many choices of coordinates are ambiguous and not locally defined! In such coordinates to determine where \emph{there} is one often needs knowledge of the geometry throughout space.  Newtonian-like gauges are like this: the coordinate chart requires determination of a gravitational potential which requires information about matter everywhere.  There exist more locally defined coordinate systems such as synchronous gauge which follow geodesics.

If all the matter is irrotational dust then Lagrangian coordinates of space-time fluid is the same a synchronous spatial coordinates which follow the dust. However Lagrangian coordinates of space-time fluid can be defined for space-times containing any type of normal matter and follow the center-of-momentum flow of that matter.  Thus a resolution of the here/there problem is to say if matter moves from \emph{here} and \emph{now} to \emph{there} and \emph{then} then \emph{here}$\,=\,$\emph{there}.

\section{Is Generation of Curvature a Gravitational Effect?}
\label{CommentsOnKurvature}

The kurvature, $K$ or $\bar{K}$, defined in eq.~\ref{Kurvature} is the difference of gravitational energy term $\propto G\times\mathrm{density}$ and a kinetic energy term $\propto\mathrm{expansion}^2$.  This difference is interpreted as a measure of specific gravitational binding energy and is clearly "gravitational" . Furthermore $\bar{K}$ is locally defined by the Ricci tensor and its derivatives and it is in this sense a purely geometrical / gravitational quantity. As I shall now show the evolution of this gravitational quantity shows little evidence of the effects of gravity. This raises the question as to "how gravitational" the evolution $K$ or $\bar{K}$ (what I have called the "generation of curvature") is.

The evolution of kurvature is derived by combining the Raychaudhuri equation with the component of contracted Bianchi identities corresponding to local conservation of energy in the center of momentum frame.  The result is given in eq.~\ref{KurvatureDensityEquationOfMotion} which may also be written
\begin{align}
    \dot{K}+\frac{2}{3}\,\theta\,K&=
    \frac{2}{9}\,\theta\,
    (2\,\sigma^2-2\,\omega^2-\vec{\nabla}\cdot\vec{a}_\mathrm{ng})
\nonumber \\
    \dot{\bar{K}}+\frac{2}{3}\,\bar{\theta}\,\bar{K}&=
    \frac{2}{9}\,\bar{\theta}\,
    (2\,\bar{\sigma}^2-2\,\bar{\omega}^2
    -{\dot{\bar{u}}^\alpha}_{;\alpha})
    -\frac{8\pi\,G\,\bar{\mu}}{3c^2}\,\bar{\sigma}^2 \ .
\label{KurvatureEquationOfMotion}
\end{align}
(the first is the classical analog of the relativistic second). These two equations are quite similar except in the relativistic case there is an additional (viscosity) term which gives the alignment of shear and anisotropic stress.  The anisotropic stress unlike the other terms is independent of the fluid velocity.  The $G$ prefactor in viscosity indicates that anisotropic stress "gravitates" in a way that affects the evolution of $\bar{K}$.  Other than viscosity the $G$ is absent from eq.~\ref{KurvatureEquationOfMotion} which suggests that all the other terms are non-gravitational.

The quantities in the space-time fluid decomposition of eq.~\ref{StressEnergyDecomposition} can be classified into \emph{kinematical}, \emph{matter} and \emph{dynamical}.  Kinematical quantities include the (center-of-momentum) velocity and its spatial derivatives in the center-of-momentum frame. These include the rate of expansion, shear and rotation.  Matter quantities include density, pressure and anisotropic stress.  \emph{Non-gravitational dynamical} quantities are the proper (non-gravitational) acceleration. \emph{Gravitational dynamical} quantities are $G\,\times$ the matter quantities. For example $G\,\rho$ gives the gravitational acceleration in Newtonian gravity. 

Clearly the viscosity term in eq.~\ref{KurvatureEquationOfMotion} depends on a gravitational dynamical quantity. Other than that the terms which govern the evolution of kurvature are either kinematical or non-gravitational dynamical. Neglecting viscosity what is missing is any gravitational dynamical terms.  This is the sense in which the generation of curvature is non-gravitational.

In contrast the Raychaudhuri equation is considered a gravitational focusing equation because it includes the gravitational focusing of streamlines through the $-4\pi\,G\,\rho$ contribution to $\dot{\theta}$ or the  $-4\pi\,G\,u^\alpha\,u^\beta\,R_{\alpha\beta}$ contribution to $\dot{\bar{\theta}}$.  The Raychaudhuri equation holds for any velocity field while the kurvature evolution equation is specific to the center-of-momentum velocity. This choice allows one to combine the Raychaudhuri equation with a contracted Bianchi identity to eliminate the gravitational focusing term.  This is what makes the evolution of kurvature simpler than the evolution of the rate of expansion and why it is "less gravitational".

Consider the line-integral solution to eq.~\ref{KurvatureEquationOfMotion}. Along any streamline, if one fixes the kinematical and dynamical terms, solutions have an undetermined constant of integration corresponding to adding constant to the curvature ($a^2\,K$ or $\bar{a}^2\bar{K}$).  If there is no qualitative difference by adding a constant then there is no qualitative difference between the evolution of fluid elements positive or negative kurvature, i.e.  between bound and unbound fluid elements which have opposite signs of kurvature.  This is a manifestation of non-gravitational nature of the evolution of curvature.

The reason why kurvature is useful for evolution of cosmological large scale structure is because its evolution is independent of gravitational dynamics.  Nearly all quantities in eq.~\ref{KurvatureEquationOfMotion} which determine the evolution of kurvature; shear, vorticity, viscosity and non-gravitational acceleration; are generally only important on small scales, below the sound horizon. The only other quantity is the rate of expansion which is important on large scales and is {\it indirectly} tied to gravitational dynamics through the Raychaudhuri equation.  However the rate of expansion  only serves to set the rate of change of kurvature and is easily mapped onto the observable cosmological redshift. It is the fact that evolution of kurvature as a function of redshift is independent of gravitational dynamics which makes it easy to track on large scales.  This simple evolution on scales larger than the sound horizon is why kurvature, either, $\bar{K}$ or $\bar{\mathcal{R}}$, is useful.

\section{Shear Happens}
\label{Shear}

The equations up until now have all been exact expressions for both a general self-gravitating (classical) fluid in Newtonian physics or a general space-times containing normal matter with occasional restrictions.  These extremely simple equations are nearly identical for the classical fluid and the space-time fluid. These equations do not provide a complete description of these systems but are sufficient to address issues related to generation of curvature which is analyzed above. Here I refine the predictions for inhomogeneities in our universe using approximations and assumptions.  The equations that follow are not exact and only meant to be rough estimations.  A more detailed analysis is given in subsequent papers.

To make these estimations one should try to restrict to "realistic" models of our universe.  What is meant by this is something along the lines of the currently popular narrative of cosmological history with sequential epochs of inflation, reheating, radiation domination, matter domination and then dark energy domination. In the simplest models the stress-energy is dominated by a single scalar field during inflation and reheating.  Such a field always has zero anisotropic stress and hence zero viscosity. At the transition between reheating and radiation domination the inflaton becomes sub-dominant the radiation fluid starts to dominate.  At this point the deviations from FLRW spacetime are small which allows one to decompose inhomogeneities into scalar, vector and tensor inhomogeneities.  In the standard narrative scalar modes dominate over vector modes, meaning $\bar{\omega}\ll\bar{\sigma}$.  During most of the radiation era the matter is highly collisional which keeps the anisotropic stress and hence the viscosity small.  During the matter era both the pressure and anisotropic stress are small which again means the vorticity is small. The same is true during dark energy domination.  In the default model we therefore expect vorticity and viscosity to give a negligible contribution to generation of curvature.  Thus the curvature density averaged over some super-horizon volumes is approximately
\begin{equation}
\overline{\Delta\bar{\rho}}[z]
\approx\frac{(1+z)^2}{(1+z_\mathrm{i})^2}\,\overline{\Delta\bar{\rho}[z_\mathrm{i}]}
+\frac{(1+z)^2}{2\pi\,G}\int_z^{z_\mathrm{i}} dz'\,
  \frac{\overline{\bar{\sigma}^2[z']}}{(1+z')^3}
\label{KurvatureDensityApproximation}
\end{equation}
The 1st term is from linear theory.  The 2nd term is generated kurvature from non-linearities as described above.  These non-linearities are inhomogeneous and stochastic leading to an uncorrelated or white noise (because of causality) spectrum on large scales.  Both scalar and tensor modes may contribute significantly to this white noise.  White noise from minimal models of density inhomogeneities has been studied in \cite{Barenboim:2025ccc,Barenboim:2025jdg}. Observational limits on large scale white noise constrain significantly constrain both scalar and tensor inhomogeneities in cosmological models.

\newpage

\backmatter

\bmhead{Acknowledgments}
The author thanks Gabriela Barenboim and Scott Dodelson for comments, criticisms and corrections. This manuscript has been authored by FermiForward Discovery Group, LLC under Contract No.~89243024CSC000002 with the U.S. Department of Energy, Office of Science, Office of High Energy Physics.

\bmhead{Author Contributions}
The entire manuscript is due solely to the author.

\bmhead{Data Availability} 
No datasets were generated or analysed during the current study.

\bmhead{Competing interests} 
The author declares no competing interests.

\bibliography{SpaceTimeFluid}

@article{ZELDOVICH1965,
  author = {Ya.~B.~Zel'dovich},
  title = {Survey of Modern Cosmology},
  jourmal = {Advances in Astronomy and Astrophysics},
  publisher = {Elsevier},
  ISSN = {0065-2180},
  volume = {3},
  year = {1965},
  pages = {241–379}
}

@book{Hawking:1973uf,
    author = {S.~W.~Hawking and G.~F.~R.~Ellis},
    title = {The Large Scale Structure of Space-Time},
    publisher = {Cambridge University Press},
    series = {Cambridge Monographs on Mathematical Physics},
    month = {2},
    year = {2023}
}

@ARTICLE{1974A&A....32..197P,
       author = {P.~J.~E.~Peebles},
        title = {The Nature of the Distribution of Galaxies},
      journal = {Astronomy \& Astrophysics},
         year = {1974},
        month = may,
       volume = {32},
        pages = {197}
}

@ARTICLE{1983ApJ...268....1C,
       author = {B.~J.~Carr  and J.~Silk},
        title = "{Can graininess in the early universe make galaxies?}",
      journal = {The Astrophysical Journal},
         year = 1983,
        month = may,
       volume = {268},
        pages = {1-16}
}

@ARTICLE{1986ApJ...302...39A,
       author = {L.~F.~Abbott and J.~Traschen},
        title = "{Causality Constraints on Cosmological Perturbations}",
      journal = {The Astrophysical Journal},
         year = 1986,
        month = mar,
       volume = {302},
        pages = {39}
}

@book{Landau1987Fluid,
  author    = {L. D.~Landau and E. M.~Lifshitz},
  title     = {Fluid Mechanics},
  series    = {Course of Theoretical Physics, Volume 6},
  edition   = {2nd},
  publisher = {Butterworth-Heinemann},
  year      = {1987},
  isbn      = {0-08-033933-6},
  }

@article{Veeraraghavan1990,
  title = {Causal compensated perturbations in cosmology},
  volume = {365},
  ISSN = {1538-4357},
  journal = {The Astrophysical Journal},
  publisher = {American Astronomical Society},
  author = {S.~Veeraraghavan and A.~Stebbins},
  year = {1990},
  month = dec,
  pages = {37}
}

@article{PhysRevD.68.103515,
  title = {Conserved cosmological perturbations},
  author = {D. H.~Lyth and D.~Wands},
  journal = {Phys. Rev. D},
  volume = {68},
  issue = {10},
  pages = {103515},
  numpages = {10},
  year = {2003},
  month = {Nov},
  publisher = {American Physical Society}
}

@article{Barenboim:2025ccc,
    author = {G.~Barenboim, A.~Ireland and A.~Stebbins},
    title = "{Large Scale White Noise and Cosmology}",
    eprint = "2511.13866",
    archivePrefix = "arXiv",
    primaryClass = "astro-ph.CO",
    preprint =  "arXiv:2511.13866",
    reportNumber = "FERMILAB-PUB-25-0832-T",
    month = "11",
    year = "2025"
}

@article{Barenboim:2025jdg,
    author = "G.~Barenboim, A.~Ireland and A.~Stebbins",
    title = "{The Noisy Universe}",
    eprint = "2511.15803",
    archivePrefix = "arXiv",
    primaryClass = "astro-ph.CO",
    reportNumber = "FERMILAB-PUB-25-0855-PPD",
    month = "11",
    year = "2025"
}

@ARTICLE{1990MNRAS.243..509E,
       author = {G.~F.~R.~Ellis},
        title = "{The evolution of inhomogeneities in expanding Newtonian cosmologies}",
      journal = {MNRAS},
         year = 1990,
        month = apr,
       volume = {243},
        pages = {509-516}
}

@article{Ellis:1994sm,
    author = "G.~F.~R.~Ellis",
    editor = "N.~S{\'a}nchez and A.~Zichichi",
    title = "{The Covariant and gauge invariant approach to perturbations in cosmology}",
    journal = "NATO Sci. Ser. C",
    volume = "467",
    pages = "1--37",
    year = "1995"
}

@article{Stebbins:2026:beta,
    author = {A.~Stebbins},
    title = "{The Space-Time Fluid in Gauge Invariant Perturbation Variables}",
    reportNumber = "FERMILAB-PUB-25-0832-T",
    year = "2026"
}

\end{document}